\documentstyle[12pt,epsf]{article}
\newcommand{\abstracts}[1]{{
\centering{\begin{minipage}{12.2truecm}
\normalsize\baselineskip=15pt
\centerline{\footnotesize ABSTRACT}\vspace*{0.3cm}
\parindent=20pt #1
\end{minipage}}\par}}
\newcommand{\diff}{\partial}
\newcommand{\dd}{\mbox{\rm d}}
\newcommand{\dD}{{\cal D}}
\newcommand{\beqn}{\begin{eqnarray}}
\newcommand{\eeqn}{\end{eqnarray}}
\newcommand{\eq}[1]{(\ref{#1})}
\newcommand{\beq}{\begin{equation}}
\newcommand{\eeq}{\end{equation}}
\newcommand{\cZ}{{\cal Z}}
\newcommand{\cO}{{\cal O}}
\newcommand{\cA}{{\cal A}}

\newcommand{\thnx}[1]{\renewcommand{\thefootnote}{\fnsymbol{footnote}}
\footnote{#1} \renewcommand{\thefootnote}{\arabic{footnote}}}

\begin{document}
\begin{center}
\vspace{-1cm}
\begin{flushright}
{\large KANAZAWA-97-04}
\end{flushright}
\vspace{1.5cm}
{\baselineskip=24pt
{\Large \bf Abelian Projection of $SU(2)$ Gluodynamics and Monopole
Condensate\thnx{talk given at the International Workshop
``Recent Developments in QCD and Hadron Physics'', 16-18 December, 1996,
Yukawa Institute, Kyoto, Japan.}}\\

\baselineskip=18pt

\vspace{1cm}

{\large
M.N.~Chernodub$^{a,b}$, M.I.~Polikarpov$^{a,b}$ and A.I.~Veselov$^a$}\\

\vspace{.5cm}
{ \it

\vspace{0.3cm}

$^a$ ITEP, B.Cheremushkinskaya 25, Moscow, 117259, Russia

\vspace{0.3cm}

$^b$ Department of Physics, Kanazawa University,\\
Kanazawa 920-11, Japan

}
}
\end{center}

\vspace{1.5cm}

\abstracts{The general properties of the abelian projection are reviewed.
We derive the explicit expression  for the abelian functional integral for
$U(1)$ abelian theory which corresponds to the abelian projection of the
$SU(2)$ gluodynamics. The numerical results for the temperature dependence of
the monopole condensate are presented.
}


\section{Introduction}
\baselineskip=14pt

The monopole mechanism \cite{tH76Ma76} of the colour confinement is
generally accepted by the lattice community. Still there are many
open questions. In order to discuss the abelian monopoles in the
vacuum of gluodynamics we have to perform the abelian
projection~\cite{tH81}. Some general properties of abelian projection
are presented in Section~2. To simplify the formulae we discuss the
Maximal Abelian (MaA) projection of the $SU(2)$ gluodynamics. The
generalization to the $SU(N)$ ($N>2$) gauge theory and to other
abelian projections is straightforward. To prove that the vacuum of
gluodynamics behave as the dual superconductor we have to show that
in the confinement phase there exists the monopole condensate. This
is a nonperturbative problem and we perform the computer simulations
of the lattice gluodynamics. To study the monopole condensate we need
the explicit expression for the operator $\Phi_{mon}(x)$, which
creates the abelian monopole at the point $x$. The operator
$\Phi_{mon}(x)$ was found for the compact electrodynamics with the
Villain form of the action by Fr\"ohlich and Marchetti \cite{FrMa87},
and it was studied numerically in Refs.\cite{Wiese}. The
generalization of the operator $\Phi_{mon}$ to the abelian projection
of the gluodynamics is suggested in Refs.~\cite{ChPoVe96}. The vacuum
expectation value of $<\Phi_{mon}>$ defines the value of the monopole
condensate.  In Section~3 we describe the numerical results for the
temperature dependence for thus defined monopole condensate. It
occurs that at the low temperature (confinement phase) $<\Phi_{mon}>
\neq 0$ and at the high temperature (deconfinement phase)
$<\Phi_{mon}>=0$. This result shows that in the confinement phase the
vacuum of the gluodynamics is similar to the dual superconductor.

\section{Maximal Abelian Projection}

The partition function of $SU(2)$ gauge theory is:

\beqn
\cZ = \int \dD A \, \exp\Bigl\{- \frac{1}{4} \int \dd^4 x \, F^2_{\mu\nu}[A]
\Bigr\}\,,
\label{PF}
\eeqn
where $F_{\mu\nu}$ is the field strength tensor:
$ F^a_{\mu\nu}[A] = \diff_\mu A^a_\nu - \diff_\nu A^a_\mu -
g \, \epsilon^{abc} A^b_\mu\,A^c_\nu$.

The MaA gauge is defined as the maximization of the functional

\beqn
 R[A]= - \int \dd^4 x\, \Bigl({(A^1_\mu)}^2 + {(A^2_\mu)}^2 \Bigr)\,,
\label{GFMaA}
\eeqn
over the gauge transformations,
$A_\mu \to A^\Omega_\mu = \Omega^+ A_\mu \Omega - \frac{i}{g} \,
\Omega^+ \diff_\mu \Omega$. The functional $R$ is invariant under the
$U(1)$ subgroup of the $SU(2)$ gauge group, therefore
the maximization condition fixes $SU(2)$ gauge freedom up to $U(1)$
gauge group.

The gauge fixing procedure is standard. We define the Faddeev--Popov unity:

\beqn
1=\Delta_{FP}[A;\lambda] \cdot \int \dD \Omega \, \exp \{ \lambda
R[A^\Omega]\}\,, \quad \lambda \to + \infty\,,
\label{FPU}
\eeqn
where $\Delta_{FP}$ is the Faddeev--Popov determinant. We substitute
the unity \eq{FPU} in the partition function \eq{PF},
shift the fields by the regular transformation $\Omega^+$:
$A \to A^{\Omega^+}$ and use the gauge invariance of the Haar measure,
the action and the Faddeev-Popov
determinant under the regular gauge transformations.
Thus we get the product of the volume of the gauge orbit,
$\int \dD \Omega$, and the partition function in the fixed gauge:

\beqn
\cZ_{MaA} = \int \dD A \, \exp\Bigl\{- \int
\dd^4 x \, \Bigl[\frac{1}{4}  F^2_{\mu\nu}[A] -
\lambda \Bigl({(A^1_\mu)}^2 + {(A^2_\mu)}^2 \Bigr) \Bigr] \Bigr\}\,
\Delta_{FP}[A;\lambda]\,.
\label{PFMaA}
\eeqn
In the non--degenerate case the FP determinant can be represented in
the form:

\beqn
\Delta_{FP}[A;\lambda] = Det^\frac{1}{2} M[A^{\Omega^{MaA}_r}] \,
\exp\Bigl\{ - \lambda R[A^{\Omega^{MaA}_r}]\Bigr\} + \dots\,,
\eeqn
where $\Omega^{MaA}_r = \Omega^{MaA}_r(A)$ is the
regular gauge transformation which corresponds to a global maximum of the
functional $R[A^\Omega]$,
the dots correspond to the terms which are suppressed in the limit
$\lambda \to \infty$; and

\beqn
M^{ab}_{xy}[A] = \frac{\diff^2 R(A^{\Omega(\omega)})
}{\diff \omega^a (x) \, \diff \omega^b (y)}
{\lower0.15cm\hbox{${\Biggr |}_{\omega=0}$}}\,,
\eeqn
$\Omega(\omega) = \exp\{i\omega^a T^a\}$, $T^a=\sigma \slash 2$
are the generators of the gauge group, $\sigma^a$ are the Pauli matrices.
In the limit $\lambda \to + \infty$ the region of the
integration over the fields $A$
reduces to region where the gauge fixing functional $R$ is maximal,
and therefore the partition function \eq{PFMaA} can be rewritten
as follows~\cite{ChPoVe95}:

\beq
        \cZ_{MaA} = \int \dD A \exp\{- S(A)\} Det^{\frac{1}{2}} \Bigl(
        M[A] \Bigr) \Gamma_{FMR}[A]\,,
                \label{GPF2}
\eeq
where $\Gamma_{FMR}[A]$
is a characteristic function of the Fundamental
Modular Region \cite{FMR} for the MaA projection\cite{ChPoVe95}.

After the abelian projection we get an abelian theory which
contains two dynamical variables, namely, abelian
gauge fields $\cA$ and monopole currents $j$. The origin of singular
monopole currents will be discussed later.
The explicit calculation shows \cite{KrLaScWi87} that these
currents form closed loops in the four-dimensional space.

The abelian gauge field $\cA$ is the component of the $SU(2)$ gauge
field which transforms as an abelian gauge field under the residual
$U(1)$ gauge transformations in the MaA gauge: $\cA_\mu = A^3_\mu$.
The $SU(2)\slash U(1)$ invariant definition of
the abelian fields is: $\cA_\mu={(A^{\Omega^{MaA}_r (A)}_\mu)}^3$.

The abelian monopole trajectory $j$ corresponding to the given gauge
field configuration $A$ is defined as follows. We maximize the
gauge fixing functional $R[A^\Omega]$ with respect to both regular
and singular gauge transformations $\Omega$.
The definition of the abelian monopole current is:

\beqn
  j_{\mu} (A) = \frac{g}{2 \pi} \epsilon_{\mu\nu\alpha\beta}
 \diff_{\nu} f_{\alpha\beta} ({(A^{\Omega^{MaA}_s (A)})}^3)\,. \label{j}
\eeqn
where $f_{\mu\nu}$ is the abelian field strength tensor
$f_{\mu\nu} (A) = \diff_\mu A_\nu - \diff_\nu A_\mu$.
The monopole currents \eq{j} are conserved and quantized \cite{KrLaScWi87},
and they can be represented as follows:

\beqn
   j_\mu (x) = \int\limits^T_0
   \dd s \, \frac{\diff {\bar x}_\mu (s)}{\diff s}
   \delta^{(4)}(x - {\bar x}(s))\,, \label{jT}
\eeqn
the four--vector ${\bar x}_\mu (s)$, $s \in [0,T)$ parameterizes the closed
monopole trajectory, ${\bar x}_\mu (0) = {\bar x}_\mu (T)$.

The field configuration $A$ contains the abelian monopoles if the
maximizing transformation $\Omega^{MaA}_s$ is singular.
The non--abelian field strength tensor
transforms under the singular gauge transformations as follows:

\beqn
F_{\mu\nu}[A] \to F_{\mu\nu}[A^{(\Omega)}] & = & \Omega^+
F_{\mu\nu}[A] \Omega + F^{sing}_{\mu\nu}[\Omega]\,,\nonumber\\
F^{sing}_{\mu\nu}[\Omega] & = & - i \Omega^+(x) [ \diff_{\mu} \diff_{\nu} -
\diff_{\nu} \diff_{\mu} ] \Omega(x)\,.
\eeqn

The abelian field strength tensor $f_{\mu\nu}$ in the MaA projection
can be decomposed into two parts,
$f_{\mu\nu} = f^r_{\mu\nu} + f^s_{\mu\nu}$, where $f^r_{\mu\nu}$ is the
regular part, $\epsilon_{\mu\nu\alpha\beta} \diff_\nu f^r_{\alpha\beta} = 0$,
and $f^s_{\mu\nu}$ is the singular part,
$f^s_{\mu\nu}(A) = F^{sing,3}_{\mu\nu}[A^{\Omega^{MaA}_s}]$.
Thus the monopoles appear due to the singularities of
the abelian fields strength tensor.

The partition function \eq{PFMaA} can be represented as the functional
integral over the abelian gauge field and over the abelian monopole
trajectories. The abelian monopoles in the MaA projection in the
functional integral formalism can be treated \cite{BaSa78} similarly to the
t'Hooft--Polyakov monopoles \cite{tHPo74} in the Georgi--Glashow model.

We define the unity

\beqn
1 & = & \Delta_{ab} [A] \cdot \sum^{+ \infty}_{n=0} \frac{1}{n!} \int
 \prod^n_{i=1}\int \dD {\bar x}^{(i)}\,
 \int \dD \cA \cdot \nonumber \\
  & & \delta \Bigl( j_{\mu} - \frac{g}{2 \pi} \epsilon_{\mu\nu\alpha\beta}
  \diff_{\nu} f_{\alpha\beta} ({(A^{\Omega^{MaA}_s (A)})}^3)\Bigr)
  \, \cdot \delta \Bigl( \cA_\mu - {(A^{\Omega^{MaA}_r (A)}_\mu)}^3 \Bigr)\,,
\label{j-unity}
\eeqn
where $\Delta_{ab} [A]$ is a ``Jacobian'' for the change of variables
$A \to \{\cA,j\}$ and the parameters ${\bar x}^{(i)}_\mu$
correspond to the disconnected parts of
the monopole trajectories. Due to the closeness of the currents $j$
the measure $\dD{\bar x}$ includes the integration
$\int \dd T \slash T$, where $T$ is defined in eq.\eq{jT},
see Ref.~\cite{BaSa78} for details.
Substituting eq.\eq{j-unity} into the partition function \eq{PF} and
integrating over the $SU(2)$ field $A$ we get:

\beqn
\cZ = \sum^{+ \infty}_{n=0} \frac{1}{n!} \int
\prod^n_{i=1}\int \dD {\bar x}^{(i)}\,
\int \dD \cA \, \exp\{- S_{U(1)}(\cA, j(s))\}\,,
\eeqn
where the abelian action is defined as follows:

\beqn
\exp\{- S_{U(1)} (\cA, j)\} =
\int \dD A \, \exp\Bigl\{- \int
\dd^4 x \, \Bigl[\frac{1}{4}  F^2_{\mu\nu}[A] -
\lambda \Bigr( {(A^1_\mu)}^2 + {(A^2_\mu)}^2 \Bigr)\Bigr]\Bigr\}\,
\nonumber\\
\Delta_{ab} [A] \, \Delta_{FP}[A;\lambda] \,
\delta \Bigl( j_{\mu} - \frac{g}{2 \pi} \epsilon_{\mu\nu\alpha\beta}
\diff_{\nu} f_{\alpha\beta} ({(A^{\Omega^{MaA}_s (A)})}^3)\Bigr)\,
\cdot \delta \Bigl( \cA_\mu - {(A^{\Omega^{MaA}_r (A)}_\mu)}^3 \Bigr)\,,
\label{AbAct}
\eeqn
here the limit $\lambda \to \infty$ is assumed.
Thus the partition function for the $SU(2)$ gluodynamics in the MaA
projection is rewritten as the partition function of some abelian
theory which contains the gauge field and the monopoles.

Usually in the abelian projection the $U(1)$ gauge invariant quantities
($\cO$) are
considered. Below we derive the explicit expression for the
 $SU(2)$ invariant quantity $\tilde \cO$ which corresponds to $\cO$.
The expectation value for the quantity $\cO$ in the MaA gauge \eq{GFC} is:

\beqn
{<\cO>}_{MaA} =
\frac{1}{\cZ_{MaA}} \int \dD A \, \exp\{ - S(A) + \lambda R[A]\} \,
\Delta_{FP}[A;\lambda] \, \cO(A)\,.
\label{E1}
\eeqn
Shifting the fields $U \to U^{\Omega}$ and integrating over $\Omega$
both in the nominator and in the denominator of expression \eq{E1} we get:

\beqn
{<\cO>}_{MaA} = <{\tilde \cO}>\,,\quad
{\tilde \cO}(A) = \frac{\int \dD \Omega \,
\exp\{ \lambda R[A^\Omega] \} \, \cO(A^\Omega)}{\int \dD \Omega \,
\exp\{ \lambda R[A^\Omega] \}}\,,
\label{tO0}
\eeqn
${\tilde \cO}$ is the $SU(2)$ invariant operator.  In the limit
$\lambda\to + \infty$ we can use the saddle point method to calculate
${\tilde \cO}$:

\beqn
{\tilde \cO}(A) = \frac{\sum\limits^{N(A)}_{j=1}
Det^\frac{1}{2} M[A^{\Omega^{(j)}}] \,
\cO(A^{\Omega^{(j)}})}{\sum\limits^{N(A)}_{k=1}
Det^\frac{1}{2} M[A^{\Omega^{(k)}}]}\,,
\label{tO}
\eeqn
where $\Omega^{(j)}$ are the $N$--degenerate
global maxima of the functional $R[A^\Omega]$
with respect to the regular gauge transformations $\Omega$:
$R[A^{\Omega^{(j)}}] = R[A^{\Omega^{(k)}}]$, $j,k=1,\dots,N$.
In the case of non--degenerate global maximum ($N=1$),
we get ${\tilde \cO}(A) = \cO(A^{\Omega^{(1)}})$.

\section{Numerical Results on the Lattice}

On the lattice the partition function of the $SU(2)$ gauge theory is:

\beqn
\cZ= \int \dD U \, \exp\{ - S(U)\}\,,\quad S(U)= \frac{\beta}{2} \,
\sum_P \, {\rm Tr} \, (1 - U_P)\,,\label{PFlat}
\eeqn
$\dD U$ is the Haar measure for the $SU(2)$ link fields $U$.

The lattice MaA projection is defined by the following
maximization condition:
\beqn
\max_{\{\Omega\}} R[U^\Omega]\,,\quad R[U]= \sum_l {\rm Tr}
(U_l \, \sigma^3 \, U^+_l \, \sigma^3)\,,
\label{GFC}
\eeqn
where the gauge transformed field $U$ is $U^\Omega_{x,\mu} = \Omega^+_x \,
U_{x,\mu} \, \Omega_{x+\hat\mu}$.

Below we present the results of the numerical calculations of the
monopole condensate on the lattice
$10^3\cdot 4$ with the anti--periodic boundary conditions.
To see that we have the
order parameter for the deconfinement phase transition it is
convenient to study the probability distribution of the monopole
creation operator $\Phi_{mon}$, the details of the calculation are
given in Ref.~\cite{ChPoVe96}.
We calculate the expectation value $<\delta (\Phi - \Phi_{mon}(x))>$.
The physical meaning has the quantity $V(\Phi)$, which is
defined as follows:

\beq
e^{-V(\Phi)} = <\delta (\Phi - \Phi_{mon}(x))>\,,
\eeq
$V(\Phi)$ is an effective potential for the monopole field. It
occurs, that in the confinement phase this potential is of the Higgs
type; in the deconfinement phase $V(\Phi)$ has the minimum at the
zero value of the field $\Phi$. The position of the minimum of the
potential, $\Phi_c$, corresponds to the value of the monopole
condensate. The quantity $\Phi_c$ strongly depends on the lattice
volume, and we use the extrapolation procedure to get $\Phi^{inf}_c$
which corresponds to the infinite volume. In Fig.~1 we show the
dependence of $\Phi^{inf}_c$ on the parameter $\beta$. At
$\beta=\beta_c$ the deconfinement phase transition takes place. From
this figure it is  clearly seen that the monopole condensate exists
($\Phi^{inf}_c\neq 0$) in the confinement phase. This fact is in the
agreement with the dual superconductor model of gluodynamics vacuum
\cite{tH76Ma76}.


\begin{figure}[htb]
\vspace{-4.5cm}
\centerline{\epsfxsize=.65\textwidth\epsfbox{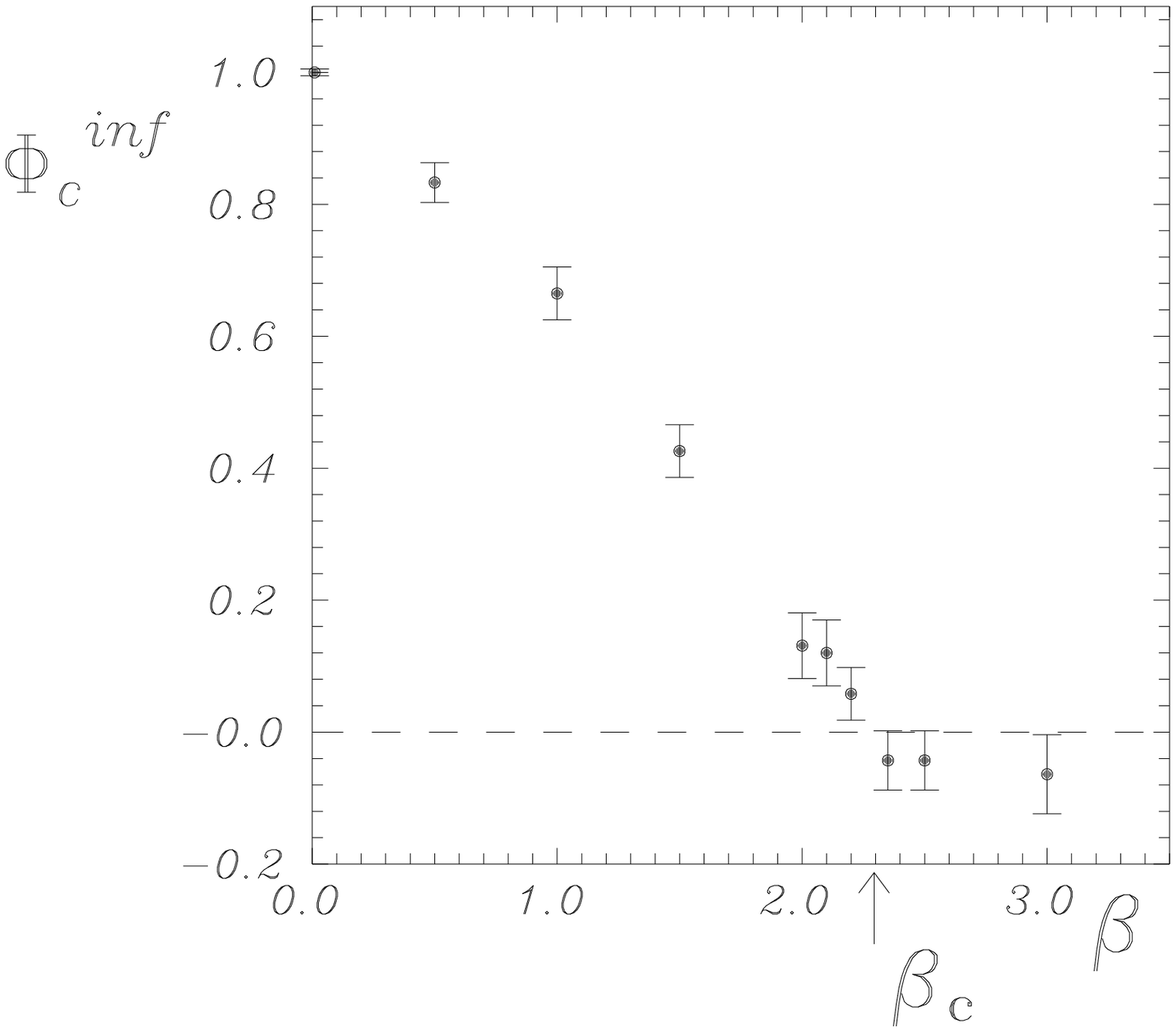}} 
\vspace{0.3cm}
\centerline{Fig.~1: The dependence of $\Phi_c^{inf}$ on $\beta$.}
\vspace{0.3cm}
\end{figure}

\section*{Acknowledgments}

M.N.Ch. and M.I.P. acknowledge the kind hospitality of the
Theoretical Department of the  Kanazawa University and they
are grateful to T.~Suzuki for interesting discussions.
Authors were supported
by the JSPS Program on Japan -- FSU scientists collaboration, by the Grants
INTAS-94-0840, INTAS--RFBR-95-0681, and by the grant 93-02-03609,
financed by the Russian Foundation for the Fundamental Sciences.

\end{document}